\documentclass[reprint,twocolumn,superscriptaddress,showpacs,nofootinbib,notitlepage,secnumarabic,amssymb, nobibnotes, aps, prd,preprintnumbers]{revtex4-1}
\usepackage[utf8]{inputenc}
\usepackage{graphicx}
\usepackage{latexsym,amsmath,amssymb,lmodern,float,url}
\usepackage{natbib}
\usepackage{color}
\usepackage{microtype}
\usepackage{bbold}
\usepackage{slashed}
\usepackage{multirow}

\usepackage{tikz}

\usepackage[colorlinks=true,backref=false, linktocpage=true,
citecolor=blue,urlcolor=blue,linkcolor=blue,pdfpagemode=UseOutlines]{hyperref}

\hypersetup{%
  bookmarksnumbered=true,
  pdftitle = {},
  pdfsubject = {},
  pdfauthor = {},
  pdfkeywords = {}
}

\renewcommand{\d}{\mathrm{d}}

\begin{document}
\preprint{FERMILAB-PUB-20-195-T}
\title{Suppressing Coherent Gauge Drift in Quantum Simulations}
\author{Henry Lamm}
\email{hlamm@fnal.gov}
\affiliation{Fermi National Accelerator Laboratory, Batavia,  Illinois, 60510, USA}
\author{Scott Lawrence}
\email{srl@umd.edu}
\affiliation{Department of Physics, University of Maryland, College Park, MD 20742, USA}
\author{Yukari Yamauchi}
\email{yyukari@umd.edu}
\affiliation{Department of Physics, University of Maryland, College Park, MD 20742, USA}
\collaboration{NuQS Collaboration}
\begin{abstract}
Simulations of field theories on noisy quantum computers must contend with errors introduced by that noise. For gauge theories, a large class of errors violate gauge symmetry, and thus may result in unphysical processes occurring in the simulation. We present a method, applicable to non-Abelian gauge theories, for suppressing coherent gauge drift errors through the repeated application of pseudorandom gauge transformation. In cases where the dominant errors are gauge-violating, we expect this method to be a practical way to improve the accuracy of NISQ-era simulations.
\end{abstract}

\maketitle

\section{Introduction}
Simulations of nonequilibrium physics of nonperturbative gauge theories, while intractable with classical resources, are possible provided sufficiently large quantum computers~\cite{Feynman:1981tf}. Whether these calculations can be achieved in the Noisy Intermediate-Scale Quantum (NISQ) era crucially depends on devising efficient algorithms that can tolerate errors from unfaithful hardware. A rich variety of error correction algorithms have been constructed (see~\cite{roffe2019quantum} for a review), but general error correction algorithms (those usable with arbitrary quantum circuits) substantially increase the qubit cost of any algorithm. It can therefore be advantageous to mitigate major sources of error in a way that exploits the properties of the calculation being performed. For gauge theories, errors which result in violations of gauge symmetry may be particularly troublesome. The origin of these errors is the difficulty in exactly matching the Hilbert space of the physical states of gauge theories to the Hilbert space of quantum computers.

Trotter-Suzuki decompositions for the time-evolution of gauge theories have been constructed that perfectly preserve gauge symmetry~\cite{Zohar:2014qma,Zohar:2016iic,Lamm:2019bik}.  When using such decompositions, gauge violations come in only through noise in the quantum hardware. Furthermore, because in many simulation schemes the vast majority of states on the quantum computer correspond to no physical state, one may expect that a substantial fraction of errors may involve the transition from a physical state to an unphysical one. This suggests that it may be profitable to suppress such errors.

In this paper we describe the suppression of coherent errors (those representable by a unitary operation) that involve drift into the unphysical sector of Hilbert space. The scheme is based on the general framework of gauge theory simulation set out in~\cite{Lamm:2019bik}, which is similar to that of~\cite{Zohar:2012ay,Zohar:2015hwa,Zohar:2016iic,Bender:2018rdp,Zohar:2018nvl,Tagliacozzo:2012df}. A wide variety of schemes for simulating gauge theories on quantum computers have been proposed~\cite{Zohar:2013zla,Hackett:2018cel,Bazavov:2015kka,Zhang:2018ufj,Unmuth-Yockey:2018xak,Unmuth-Yockey:2018ugm,Zache:2018jbt,Stryker:2018efp,Raychowdhury:2018osk,Kaplan:2018vnj,Klco:2019evd,Alexandru:2019nsa,Zohar:2012ay,Zohar:2015hwa,Zohar:2016iic,Bender:2018rdp,Zohar:2018nvl,Tagliacozzo:2012df,Lamm:2019bik,Kreshchuk:2020dla,Liu:2020eoa,Harmalkar:2020mpd}; the mitigation method discussed here should be broadly applicable to simulation methods in which the physical Hilbert space is only a small subspace of the total Hilbert space on the quantum computer. Simulation schemes that make use of gauge fixing do not have this property, and thus do not have the opportunity for gauge-violating errors, at the cost of higher complexity.

Existing gauge simulations on quantum processors have dealt with gauge-violating noise through the use of post-selection and by extrapolating results to the noiseless limit, e.g.\ \cite{Klco:2019evd}. The proposal of ~\cite{Stryker:2018efp}, formulated for Abelian theories, reduces gauge violation by repeatedly measuring Gauss's law.
A separate proposal, also for Abelian theories~\cite{Halimeh:2019svu}, creates an energy penalty for nonphysical states. The method we describe here can be viewed as a form of energy penalty, suitable for non-Abelian theories. In Section~\ref{sec:discussion}, we will show how the method can be slightly modified to correspond instead to the repeated measurement of Gauss's law.

\section{Method}
We begin with definitions and notation. In simulating a lattice gauge theory with gauge group $G$, each link has a local Hilbert space $\mathcal H_\ell = \mathbb C G$ isomorphic to the space of square-integrable functions on $G$. For simplicity, assume that $G$ is a finite group, with $|G| = 2^q$, such that an isomorphism exists between $\mathcal H_\ell$ and the Hilbert space of $q$ qubits. For a pure gauge theory (that is, neglecting matter fields), the Hilbert space of a lattice with $L$ links is given by the tensor product of $L$ local spaces, $\mathcal H = \mathcal H_\ell^{\otimes L}$. This space is isomorphic to the Hilbert space of $qL$ qubits, and we will not distinguish between the two.

A gauge transformation is given by an assignment of an element of $G$ to each of the $V$ sites of the lattice\footnote{For groups with a nontrivial center, there is a small amount of redundancy in this description, which we will ignore, so that some expressions will be off by a factor of $|Z(G)|$.}. Each transformation $g$ yields a unitary operator $\phi(g)$ acting on the Hilbert space $\mathcal H$. This Hilbert space is larger than the Hilbert space of the physical theory. The Hilbert space of the physical theory is spanned by only those states which are invariant under all gauge transformations $\phi(g)$. In other words, the physical Hilbert space is the gauge projection of the original space, $\mathcal H_P = P \mathcal H$, where the gauge projection operator is obtained by integrating over all gauge transformations:
\begin{equation}
P \propto \int \d^V\!g\; \phi(g)
\text.
\end{equation}
How large is $\mathcal H$ compared to $\mathcal H_P$? It is convenient to think in a computational basis, spanned by states labelled by an assignment of a single element of $G$ to each link; i.e., the wavefunction has support only on a single field configuration. On a $d$-dimensional lattice, there are $|G|^{dV}$ basis states, so that is the dimension of $\mathcal H$. Generically, each such state is connected via gauge transformations to $|G|^{V}$ other basis states, and only the symmetric superposition of all these basis states is physical. Thus we see that for each physical state, there is an associated Hilbert space of dimension $|G|^{V}$, of which all directions but one are nonphysical. This suggests drifting into the unphysical Hilbert space will be a large component of any noise.

The scheme of~\cite{Lamm:2019bik} defines, via the Suzuki-Trotter decomposition, a unitary operator $U$ which approximates a short amount $\delta$ of time-evolution, so that $U^{t/\delta}$ is a good approximation to $e^{-i H t}$. Note that $U$ commutes with all gauge transformations, and thus $U|\Psi\rangle$ is a physical state whenever $|\Psi\rangle \in \mathcal H_P$ was a physical state. This property is broken by the presence of errors in the quantum computer. We will consider coherent errors, by which a physical state $|\Psi\rangle$ is transformed to a state
\begin{equation}
    U_E |\Psi\rangle = 
    \sqrt{1 - \epsilon^2}|\Psi\rangle + \epsilon |\omega\rangle
\end{equation}
where $|\omega\rangle$ is an unphysical state, orthogonal to $\mathcal H_P$. We are explicitly ignoring non-coherent errors resulting from interactions with the environment; when the environment is traced out, these yield classical uncertainty in the state of the quantum processor.

It was previously proposed~\cite{Stryker:2018efp} (for Abelian theories) that these coherent errors be suppressed in a non-unitary way by repeatedly measuring Gauss's law at each site. Intuitively, these coherent errors can also be suppressed in a unitary way by modifying the Hamiltonian $H$ which defines the time evolution; this was explored for Abelian theories in~\cite{Halimeh:2019svu}, but is in principle applicable to arbitrary gauge groups. Define a Hermitian operator $H_G$ as follows: $H_G |\Psi\rangle = 0$ for any physical state $|\Psi\rangle \in \mathcal H_P$, and $H_G |\omega\rangle = 1$ for all states orthogonal to the physical subspace. Adding (a large scalar multiple of) $H_G$ thus imposes an energy penalty on transitions to unphysical states.

At the core of our method is the task of creating an effective energy penalty on nonphysical states. At this point, we may seek to write $H_G$ in terms of the fundamental fields of the lattice gauge theory. For a non-Abelian gauge group, this is a difficult task (just as a gauge oracle along the lines of~\cite{Stryker:2018efp} is difficult to construct). However, the object needed for a quantum simulation is not $H_G$, but rather the unitary $e^{-i H_G T}$ in the limit of large $T$, and this object is much easier to construct. Roughly speaking, this unitary performs a random gauge transformation. This suggests the following mitigation of coherent gauge drift errors: after every Trotterization time step, sample and perform a random gauge transformation from the Haar measure on $G^V$. Physical states are unaffected by this operation, while unphysical states have their phase rotated by a random amount, just as they would under faithful implementation of a large energy penalty.

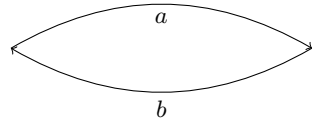
\begin{figure}
    \centering
    \begin{tikzpicture}
    \draw[->] (0,0) to[out=30,in=150] node[below] {$a$} (4,0);
    \draw[->] (4,0) to[out=-150,in=-30] node[below] {$b$}(0,0);
    \end{tikzpicture}
    \caption{A $\mathbb Z_2$ gauge theory with a single plaquette, consisting of two links.}
    \label{fig:plaquette}
\end{figure}

Let us see how this approach performs with a concrete example: a $\mathbb Z_2$ gauge theory on two links (labelled $a$ and $b$), shown in Figure~\ref{fig:plaquette} and defined by the Hamiltonian
\begin{equation}
    H = \sigma_x(a) + \sigma_x(b) + \sigma_z(a)\sigma_z(b)
    \text.
\end{equation}
The Hilbert space $\mathcal H$ has four states, spanned by (in the $z$-basis) $|00\rangle$, $|01\rangle$, $|10\rangle$, and $|11\rangle$. There is only one nontrivial gauge transformation, $\phi(g) = \sigma_x(a)\sigma_x(b)$, which has the effect of flipping both qubits. We denote the two physical states by $|0_+\rangle = |00\rangle + |11\rangle$ and $|1_+\rangle = |01\rangle + |10\rangle$. The orthogonal states are $|0_-\rangle = |00\rangle - |11\rangle$ and $|1_-\rangle = |01\rangle - |10\rangle$.

Consider the simple case of coherent gauge-drift from $|0_+\rangle$ to $|0_-\rangle$ defined by
\begin{align}
U_E =&
\left(\begin{matrix}
|0_+\rangle & |0_-\rangle
\end{matrix}\right)
\left(
\begin{matrix}
\sqrt{1-\epsilon^2} & \epsilon\\
-\epsilon & \sqrt{1-\epsilon^2}\\
\end{matrix}
\right)
\left(\begin{matrix}
\langle 0_+| \\ \langle 0_-|
\end{matrix}\right)
\nonumber\\&\hspace{0.2cm}+ |1_+\rangle\langle1_+| + |1_-\rangle\langle1_-|
\text.
\end{align}
With many applications of $U_E$, the unphysical branch of the wavefunction builds up approximately linearly over time, yielding a substantial (incorrect) contribution to the final measurement made:
\begin{equation}
    U_E U_E |0_+\rangle
    = (1 - 2 \epsilon^2) |0_+\rangle + 2 \epsilon (1 - \epsilon^2)^{1/2} |0_-\rangle
    \text.
\end{equation}
More realistically, interspersed application of the time-evolution operator $e^{-i H t}$ causes some phase cancellation. We will assume this phase cancellation to be small and ignore it; where it is not small, it is because $H$ already imposes a large energy penalty on the unphysical subspace.

What happens when we intersperse application of $U_E$ with gauge transformations? The two transformations available are the trivial transformation, which acts as the identity operator on the Hilbert space), and the nontrivial bitflip transformation $\phi(g)$, which acts to flip the signs of the states $|0_-\rangle$ and $|1_-\rangle$, while leaving the two physical states unchanged. When the nontrivial transformation is used, we find $U_E \phi(g) U_E |0_+\rangle = |0_+\rangle$. The gauge drift exactly cancels! For a more general gauge drift, an $O(\epsilon^2)$ deviation from $|0_+\rangle$ may be created.

For general gauge groups and drifts, we cannot hope to achieve exact cancellation in all cases. In the $\mathbb Z_2$ case above, when the gauge transformation performed is chosen at random, instead of being optimally selected for a particular drift operator $U_E$, the amplitude of the unphysical branch of the wavefunction performs a random walk with step size $\epsilon$. Thus, at time $t$ this amplitude is $\sim \epsilon \sqrt{t/\delta}$, an asymptotic improvement over the original linear growth in $t/\delta$. In the case where the noise $\epsilon$ scales linearly with the timestep size $\delta$, this results in the complete removal of gauge drift in the continuous-time limit of the Trotterization.

We turn now to consider the general case. An arbitrary gauge drift operator may be decomposed as
\begin{equation}\label{eq:UE}
    U_E = A + \epsilon V
\end{equation}
where $A$ is diagonal, and the only nonzero matrix elements of $V$ are those taken between the physical and unphysical subspaces. The build-up over time of gauge drift can be analyzed by expanding $U^n_E$ in $n+1$ terms, organized by the time of the first appearance of $V$. In the absence of intervening gauge transformations, we have
\begin{equation}\label{eq:expansion}
    U^n_E =
    \epsilon V U_E^{n-1}
    +
    \epsilon A V U_E^{n-2}
    + \cdots + A^n
    \text.
\end{equation}
A transition amplitude between the initial physical state $|\psi\rangle$, and any other state (whether physical or not), receives contributions from $t$ individual terms. Because $U$ is unitary, matrix elements of $U$ are bounded in magnitude by $1$, and thus each drift term in the total amplitude is bounded above by $\epsilon$. Thus, the gauge drift results in errors that are $O(\epsilon t/\delta)$. 

Introducing random gauge transformations, each term in Eq.~(\ref{eq:expansion}) receives a random phase. For a discrete gauge group, every gauge transformation $g$ must be of finite order $|g|$, and the phases are constrained to be $|g|$-roots of unity. In the case of the single-plaquette $\mathbb Z_2$ theory, the phases can be only $\pm 1$. In all cases, the phases of each term are uncorrelated with each other, and in fact are uncorrelated with the size of the term. As a result, the series constitutes a random walk with step size bounded above by $\epsilon$, and thus the growth of errors is $O(\epsilon \sqrt{t/\delta})$.

This logic suggests a specific characterization of the sorts of errors mitigated by this procedure. Errors that begin with a transition into an unphysical state, and which remain in that unphysical state for long enough that a gauge transformation can be applied, are mitigated. All other errors are unaffected by this procedure.

\begin{figure}[t]
\includegraphics[width=\linewidth]{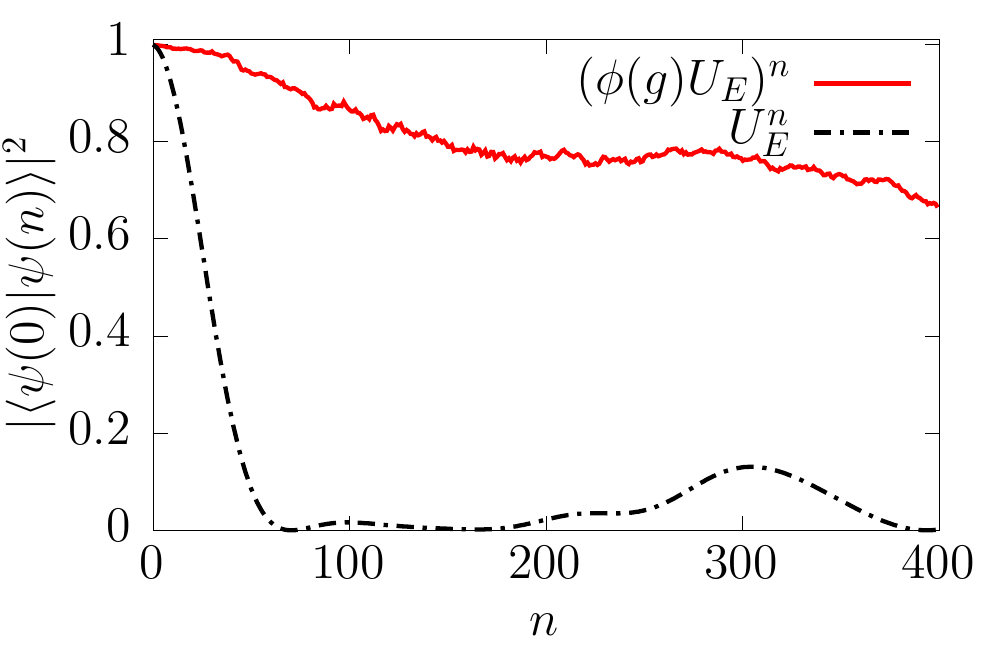}
\caption{The probability of remaining in the initial physical state $|\psi(0)\rangle$ after $n$ time steps, with (red) and without (black) intervening gauge transformations. The gauge drift operator is constructed by exponentiating the unphysical projection $H - PHP$ of a random Hermitian matrix $H$, with real and imaginary parts uniformly distributed in $[-0.01,0.01]$.\label{d3}}
\end{figure}

As a demonstration, Fig.~\ref{d3} shows the simulation of $D_3$ lattice gauge theory on the two link model shown in Fig.~\ref{fig:plaquette}. As $D_3$, the dihedral group on the triangle, is the smallest non-Abelian group, this is the simplest possible nontrivial non-Abelian gauge theory. We simulate, in the absence of a physical Hamiltonian, the effect of gauge drift with and without intervening random gauge transformations. In the absence of gauge drift, the simulation would remain forever in the initial state; the gauge drift quickly rotates the state into one approximately orthogonal to the initial state. Intervening random gauge transformations are seen to dramatically suppress the gauge drift.

\section{Performance}
How may such samplings from the gauge group be effected in practice? The notion of ``perform a random gauge transformation,'' taken literally, describes a non-unitary operation. One option is to sample gauge transformations classically ahead of time, and hard-code them into the quantum evolution circuit. This results in a circuit which contains (particularly for a large number of Trotter steps) a lot of information, which may be inconvenient. To keep the time evolution homogeneous, we may introduce a pseudorandom number generator on the quantum processor itself. For instance, we may introduce an ancillary `clock' register, which increments by $1$ with every Trotter step. A cryptographic hash function, or a random circuit as described by~\cite{Alexandru:2019dmv}, is then used to select a pseudorandom element of $G$ from the value in the clock, storing the result in an additional ancillary register. This is then repeated for each site. This technique has a drawback: after these ancillary registers are used to perform the gauge transformation, they are entangled with the state of the physical registers, and must be disentangled before they can be measured and the computation can proceed. A method for disentangling these registers is described in~\cite{lawrence_2020}.

We have assumed that the random gauge transformations are sampled fairly from the Haar measure; however, this is unnecessary to mitigate gauge-drift errors. To see this, note that a common method for sampling (approximately) from the Haar measure is to select two (or several) generic elements $g,h \in G$, and sample a random string of some fixed length, e.g.\ $ghggh$. In the limit of long strings, this will converge to the Haar measure for almost any probability distribution on $G$. Thus, if we sample unfairly from $G$ and perform multiple independent gauge rotations, we approximate the process of sampling fairly from $G$, and suppress gauge-drift exactly as discussed above. This is substantially cheaper in practice.

We can estimate the resource cost of this procedure, in terms of the costs of the underlying operations defined in~\cite{Lamm:2019bik}. A small number of ancillary qubits, including one full $G$-register, are required in order to implement the random sampling; this cost does not scale with the size of the system being simulated. The actual sampling of gauge elements is likewise quite cheap. In particular, because the correctness of the algorithm does not depend on the gauge elements being fairly sampled, a quite short random circuit of the sort described in~\cite{Alexandru:2019dmv} is sufficient. That leaves the task of performing the gauge transformation, which requires about as many $G$-multiplication gates as the original quantum simulation did. Thus, the gate cost of the simulation is approximately doubled by the inclusion of this procedure.

\section{Discussion}\label{sec:discussion}
So far we have depended on approximate cancellations of random phases to suppress the gauge drift. Another possibility, along the lines suggested by~\cite{Stryker:2018efp}, is to perform measurements of Gauss's law repeatedly during the time evolution, preventing unphysical amplitudes from building up via the quantum Zeno effect. Concretely, we introduce a single ancillary qubit beginning in state $\frac 1 {\sqrt 2}\big(|0\rangle + |1\rangle\big)$, and perform a random gauge transformation controlled on that qubit. If the initial state was mostly a physical state $|\psi\rangle$ with an amplitude of $\epsilon$ in an unphysical state $|u\rangle$, then (assuming without loss of generality the state $|u\rangle$ to be an eigenstate of the particular gauge transformation) the effect of the controlled gauge transformation is given by
\begin{align}
    \phi(g)_C \big(|\psi\rangle + &\epsilon|u\rangle\big) (|0\rangle + |1\rangle)
    =\nonumber\\
    &\big(|\psi\rangle + \epsilon|u\rangle\big) |0\rangle 
    +
    \big(|\psi\rangle + \epsilon e^{i\theta}|u\rangle\big) |1\rangle
\end{align}
where $\theta$ is the eigenvalue of the Gauss's law operator. Measuring $\sigma_x$ on the ancillary qubit, we find a probability of $\frac{\epsilon^2}{2}\cos\theta$ that we `fail', landing in the $|0\rangle - |1\rangle$ state and destroying all gauge-invariant information. The remainder of the time, the state after measurement is proportional to 
\begin{equation}
    |\psi\rangle + \epsilon \frac{1 + e^{i\theta}}{2}|u\rangle
\end{equation}
Thus the unphysical amplitude has been reduced, by an amount determined by how sensitive the unphysical state was to the particular gauge transformation performed. In the best-case scenario (common for a $\mathbb Z_2$ gauge theory) the unphysical amplitude vanishes after this procedure.

It is worth noting that the scheme described above is not limited to the simulation of gauge theories. Rather, it may be applied in any case where the Hilbert space of the quantum computer contains many states not mapped to the physically meaningful Hilbert space. This is true, for instance, in a simulation of a quantum spin chain of spin-$1$ particles, such that each spin has a three-dimensional local Hilbert space. It is natural to encode this space on a quantum computer by using two qubits for each spin; however, with $V$ spins, that results in $4^V - 3^V$ unphysical states. As with the gauge theory, at large volumes, the vast majority of states on the quantum computer have no physical meaning. Performing random transformations on the unphysical parts of Hilbert space will suppress coherent errors driving the processor into those states.

We have described a scheme by which coherent gauge-drift errors in a quantum gauge theory simulation may be consistently suppressed. By interspersing random gauge transformations, the coherent drift is transformed to a coherent walk, causing it to grow sub-linearly in time. This has the potential to reduce the impact of a large class of errors in the NISQ era. This scheme does not address errors that act directly within the physical subspace, nor does it mitigate decohering errors.

\begin{acknowledgements}
H.L.\ is supported by a Department of Energy QuantiSED grant. Fermilab is operated by Fermi Research Alliance, LLC under contract number DE-AC02-07CH11359 with the United States Department of Energy. S.L.\ and Y.Y.\ are supported by the U.S. Department of Energy under Contract No.~DE-FG02-93ER-40762.
\end{acknowledgements}

\bibliography{suppression}
\end{document}